\documentclass{emulateapj}
\usepackage{epsfig}
\usepackage{epstopdf}
\usepackage{graphicx}
\usepackage{amsmath}
\usepackage{amssymb}
\usepackage{natbib}
\newcommand{\msun}{\ensuremath{{\rm M_{\odot}}}}

\shorttitle{Minimum Galaxy Mass at High-z} 
\shortauthors{Mu{\~n}oz \& Loeb}

\begin{document}

\title{Constraining the Minimum Mass of High-Redshift Galaxies and Their Contribution to the Ionization State of the IGM}

\author{J.~A.~Mu{\~n}oz\altaffilmark{1,2} \& A.~Loeb\altaffilmark{2}}
\altaffiltext{1}{University of California Los Angeles, Department of Physics and Astronomy; Los Angeles, CA 90095, USA;}
\altaffiltext{2}{Harvard-Smithsonian Center for Astrophysics, 60 Garden
Street, Cambridge, MA 02138, USA;}

\begin{abstract}

We model the latest {\it{HST}} WFPC3/IR observations of $\ga 100$ galaxies at redshifts $z=7$--8 in terms of a hierarchical galaxy formation model with starburst activity.  Our model provides a distribution of UV luminosities per dark matter halo of a given mass and a natural explanation for the fraction of halos hosting galaxies.  The observed luminosity function is best fit with a minimum halo mass per galaxy of $10^{9.4^{+0.3}_{-0.9}}\,\msun$, corresponding to a virial temperature of $10^{4.9^{+0.2}_{-0.7}}\,{\rm K}$.  Extrapolating to faint, undetected galaxies, the total production rate of ionizing radiation depends critically on this minimum mass.  Future measurements with {\it {JWST}} should determine whether the entire galaxy population can comfortably account for the UV background required to keep the intergalactic medium ionized.

\end{abstract}


\section{Introduction}

The pursuit for the first galaxies has recently entered a new phase as observations at redshifts $z \gtrsim 6$ have now probed the epoch before cosmic reionization was complete enough to fill in the Gunn-Peterson trough.  The WFC3/IR Camera on {\it {HST}} has recently detected a sample of more than 60 $z\sim7$ and nearly 50 $z\sim8$ Lyman-break galaxies (LBGs) \citep{Bouwens10e} and provided constraints on the galaxy abundance as early as $z\sim10$ \citep{Bouwens10c}.  The UV spectral slopes of these faint sources were found to be very flat, perhaps indicating dust-free environments \citep{Bouwens10a}, and their stellar masses have been inferred from measurements in the rest-frame optical \citep{Gonzalez10, Labbe10a, Labbe10b}.  These observations inform theoretical models of galaxy formation and attempt to probe the amount of radiation available to affect the ionization state of the intergalactic medium (IGM) but are often limited by survey sensitivity.  The traditional interpretation of such data relies on an assumed ratio of UV luminosity to star formation rate (SFR) that requires burst ages longer than the exponential burst time-scale and time-scales longer than 1 Gyr \citep{Madau98}.  This assumption cannot be satisfied at redshifts $z>6$, where the age of the Universe is shorter than 1 Gyr.

Many theoretical studies based on numerical and semi-analytic techniques, have shown that the luminosity function (LF) of high-redshift LBGs can be explained by the hierarchical formation of dark matter halos whose associated baryonic gas forms stars in merger-generated bursts \citep[e.g.][]{Baugh05, Finlator10, Lacey10, Salvaterra10}.  Analytic work has tried to fit simpler models to the observed LF to probe the duty cycle and mass-to-light ratio of observed galaxies \citep[e.g]{SLE07, Trenti10} assigning a single galaxy luminosity for each halo mass, while others have considered that the mass of a host halo may merely define the probability distribution from which a galaxy's luminosity is drawn \citep[e.g.][]{CM05a, CM05b, CO06, VO04, VO06, VO08}.  Associating the mass of underlying halos with observed luminosity is crucial for describing the clustering properties and bias of high-redshift galaxies as well as the contribution of cosmic variance to fluctuations in the measured abundance from field-to-field \citep[e.g.][]{TS08, ML08a, Overzier09, Munoz10, Robertson10a, Robertson10b}.  The relationship may also provide insights into how much ionizing radiation is provided by galaxies too faint to be detected with current instruments.  

The amount of currently undetected UV radiation at high redshifts is unknown.  While the Early Release Science observations with WFC3/IR can probe down to ~27.5 AB mag \citep{Bouwens10e}, it is almost certain that many fainter galaxies remain to be observed by {\it {JWST}} \citep[e.g.][]{BL00b, WL06, SF06}.  Galaxies should exist in halos down to a mass below which the assembly or retainment of gas is suppressed.  While these dwarf galaxies near the suppression limit may be faint, their abundance may make them, in aggregate, large contributors to the total UV background.  The exact suppression mass of galaxy formation is due to an unknown combination of the heating of the IGM during reionization and thermal and mechanical feedback by internal mechanisms such as supernovae \citep{WL06}.

We propose that the suppression of galaxy formation below a certain halo mass threshold may already be evident from the currently observed LFs at $z \gtrsim 6$.  Because bright galaxies are formed primarily through mergers among fainter ones and since the contribution to the population of bright galaxies from lower mass halos shining at the bright end of their luminosity distributions need not be negligible, we expect the decrease in the faint end of the LF due to the suppression of galaxy formation to be gradual and extend to larger luminosities than previously anticipated.  We couple hierarchical merger trees to a simple model of star formation to calculate the luminosity distribution function (LDF) for galaxies as a function of their host halo mass and the resulting LF.  The suppression mass of galaxy formation applied to the variety of merger histories provides us with a physically motivated explanation for the fraction of halos that are forming stars at the time of observation (i.e. the duty cycle) as a function of mass.  We calculate the amount of unobserved UV radiation at $z=6$, 7, and 8 and test the applicability of the \citet{Madau98} relationship between UV luminosity and SFR.  We note that \citet{Raicevic10} recently considered the effect of photoionization on the LF using the GALFORM model of galaxy formation \citep{Cole00, Baugh05}, but our models produce very different results at the faintest luminosities.  This is likely because photoionization in the GALFORM model only affects gas {\it {cooling}} while allowing already cold gas to form stars even in the smallest halos.

In \S\ref{sec:model}, we describe the merger trees, star formation model, and suppression prescriptions used to populate the LDF for each halo mass.  We describe the resulting shape and mass-dependence of the LDF in \S\ref{sec:Ldist} and discuss the physical origin of the luminous duty cycle of halos in \S\ref{sec:eDC}.  In \S\ref{sec:LF}, we fit the mean of the LDFs, the resulting star formation efficiency, and the mass at which star formation is suppressed to match the observed LFs at $z \gtrsim 6$.  In \S\ref{sec:sfr} we compare the star formation rate to the prediction from the UV luminosity.  We then discuss the implications for the ionization state of the IGM of the suppressed star formation in low-mass halos and the abundance of faint galaxies yet to be observed in \S\ref{sec:IGM}.  Finally, we summarize our main conclusions in \S\ref{sec:conc}.

\section{The Model}\label{sec:model}

Our calculation of the LDF has two main components: a merger tree builder and a star formation model.  Throughout, we assume a flat, $\Lambda$CDM cosmology with cosmological parameters taken from the {\it {WMAP-5}} data release \citep{Komatsu09, Dunkley09}.

\subsection{Merger Histories}\label{sec:mergers}

We generate merger trees based on the extended Press-Schechter procedure outlined in \citet{SK99}.  The method selects, for each descendent, a series of progenitors from the mass-weighted conditional mass function truncated at an upper mass limit for each subsequently selected progenitor such that the total mass in progenitors does not exceed the descendent mass.  Once the difference in mass between the descendent and the growing list of progenitors falls below $M_{\rm res}$, the resolution limit of the algorithm, the remaining mass is assigned as diffuse accretion.  If a descendent has two or more progenitors, we determine the merger ratio by considering the two largest progenitors.  If the mass ratio between the two largest progenitors is smaller than 1:3, we denote the interaction to be a major merger; all other configurations are minor mergers.  We have tested our procedure using a threshold ratio of 1:7 and found no noticeable difference in our results.  The algorithm is then iterated on each progenitor until the tree is ended when the masses of all progenitors have fallen below $M_{\rm res}$. 

\subsection{Starbursts}\label{sec:bursts}

Each halo in the merger tree is assumed initially to contain an amount of baryonic gas equal to a fixed fraction $(\Omega_{\rm b}/\Omega_{\rm m})$ of its halo mass.  This gas is gradually converted into stars through bursts of star formation.  There is a great body of evidence that the starbursts that illuminate LBGs are generated in mergers rather than in a quiescent mode \citep[e.g.][]{Baugh05, Lacey10}.  Therefore, we ignore all quiescent star formation and generate starbursts exclusively during major mergers. 

After a minor merger, all starbursts taking place in associated branches of the tree are allowed to continue simultaneously with their own reservoirs of gas.  If not fully coalesced at the time of observation, these simultaneous bursts may appears as multiple cores in the galaxy morphology or simply be beyond our current ability to resolve \citep{Oesch10b}.  However, in a major merger, all of these bursts are shut off and a new one is begun.  The gas remaining from all progenitors is assumed to be instantaneously funneled to the center where it forms a new disk.  Following \citet{MMW98} and \citet{BL00a}, we assume the disk to have an exponential shape such that the surface density falls off as $\Sigma=\Sigma_0\,e^{-r/R_{\rm d}}$.  At high redshift, when the energy density of the universe is dominated by the contribution from matter, the corresponding exponential size scale of the disk is given by 
\begin{eqnarray}\label{eq:Rd} 
R_{\rm d}&=&\frac{1}{\sqrt{2}}\,\left(\frac{j_{\rm d}}{m_{\rm d}}\right)\,\lambda\,r_{\rm vir} \nonumber \\
&\approx&0.1\,h^{-1}\,{\rm kpc} \left(\frac{\lambda}{0.05}\right) \left(\frac{j_{\rm d}}{m_{\rm d}}\right) \left(\frac{v_{\rm c}}{30\,{\rm km/s}}\right) \left[\frac{H(z\!=\!7)}{H_0}\right]^{-1},
\end{eqnarray}
where we take the specific angular momentum of the disk to be equal to that of the halo (i.e. $j_{\rm d}/m_{\rm d}=1$), $r_{\rm vir}$ is the halo virial radius, $v_{\rm c}$ is the circular velocity of the halo, $H$ is the Hubble parameter, and $\lambda$ is the spin parameter which we draw randomly for each disk from a log-normal distribution centered at $\bar{\lambda}=0.05$ with a standard deviation $\sigma_{\lambda}=0.5$ in log-space.  The central surface density is $\Sigma_0=M_{\rm gas}/2\,\pi\,R_{\rm d}^2$.

Following \citet{Kennicutt98}, the surface star formation rate density is given by $\Sigma_{\rm SFR}=\epsilon\,\Sigma_{\rm gas}/t_{\rm dyn}$, where $\epsilon$ is the fraction of gas converted to stars in a dynamical time $t_{\rm dyn}$. We verified that the disks under consideration are unstable to fragmention (i.e. have a Toomre $Q$-parameter smaller than unity).  Using surface densities averaged over the exponential scale radius of the disk and $t_{\rm dyn}=2\,\pi\,R_{\rm d}/v_{\rm c}$, $\epsilon$ is found empirically to be $\sim 20\%$.  However, this relation also holds in azimuthally-averaged rings at radius $r$ with $t_{\rm dyn}=2\,\pi\,r/v_{\rm c}$.  Since we are interested in the total star formation rate produced entire disk from gas added at any radius, we integrate through the disk considering separately the contributions inside and outside the exponential scale radius.
\begin{eqnarray}\label{eq:Msfr}
\dot{M}_{\star}&=&\dot{M}_{\star}(<R_{\rm d})+2\,\pi\,\int_{R_{\rm d}}^{\infty}\!\! r\,\Sigma_{\rm SFR}(r)\,dr \nonumber \\
&=&\epsilon\,\Sigma_0\,v_{\rm c}\,\left[\left(\int_{R_{\rm d}}^{\infty}\!\! \frac{r\,e^{-r/R_{\rm d}}}{R_{\rm d}}\,dr\right)+\left(\int_{R_{\rm d}}^{\infty}\!\! e^{-r/R_{\rm d}}\,dr\right)\right] \nonumber \\
&=&\epsilon\,\Sigma_0\,v_{\rm c}\,R_{\rm d}\,\left[\left(e^{-1}\right)+\left(1-2\,e^{-1}\right)\right]
\end{eqnarray}
Substituting for $R_{\rm d}$ and $\Sigma_0$, we find
\begin{eqnarray}\label{eq:Msfr7}
\dot{M}_{\star}&\approx&0.66\,{\rm \msun\,/yr} \nonumber \\
&&\times \left(\frac{M_{\rm gas}}{10^8\,h^{-1}\,\msun}\right) \left(\frac{\epsilon}{0.2}\right) \left(\frac{\lambda}{0.05}\right)^{-1} \!\left[\frac{H}{H(z\!=\!7)}\right]
\end{eqnarray}
Since $\dot{M}_{\star}=-\dot{M}_{\rm gas}$ in a single burst between major mergers, equation (\ref{eq:Msfr7}) represents a differential equation in $M_{\star}$ (or $M_{\rm gas}$) whose solution is an exponential with a time scale given by:
\begin{equation}\label{eq:tau7}
\tau \approx 0.27\,\,t_{\rm age}(z) \left(\frac{\epsilon}{0.2}\right)^{-1} \left(\frac{\lambda}{0.05}\right), 
\end{equation}
where 
\begin{equation}\label{eq:tau7}
t_{\rm age}(z) \approx 0.52 \,h^{-1}\,{\rm Gyr}\,\left(\frac{1+z}{8}\right)^{-3/2}
\end{equation}
is the age of the universe at redshift $z$.

In each time step, we use results from a simple stellar population generated by Starburst99 \citep{Leitherer99} to calculate the contribution of each newly added group of stars to the final luminosity at 1500 \AA.  We assume a Salpeter initial mass function (IMF) with a slope of 2.35 between 1 and $100\,\msun$ and a metallicity 4\% of solar.

Finally, we include the effect of a suppression in galaxy formation below halos of a given mass.  Some combination of processes, such as supernovae feedback or photoionization, may push or heat the gas so that it escapes the gravitational potential well of the halo.  Thus, no starbursts will be generated in halos smaller than $M_{\rm supp}$, and neither can a halo smaller than $M_{\rm supp}$ be a constituent in a starburst-generating major merger even if the resulting halo is larger than $M_{\rm supp}$.  These two conditions (but especially the latter) combine to make inactive even some halos larger than $M_{\rm supp}$ (see \S\ref{sec:eDC}).  The simplest way to incorporate these effects into our model is simply to set $M_{\rm res}=M_{\rm supp}$ in generating the merger trees.  This prescription is not quite realistic, since the feedback processes that suppress star formation are undoubtedly time dependent, especially during reionization.  However, for simplicity, we assume that the contributions to the LF from minihalos smaller than $M_{\rm supp}$ before reionization are minimal by the redshifts considered here.  For convenience, we define 
\begin{equation}\label{eq:msupp}
m_{\rm h} \equiv {\rm log_{10}}\,(M_{\rm h}/M_{\odot}), \nonumber
\end{equation}
\begin{equation}
m_{\rm supp} \equiv {\rm log_{10}}\,(M_{\rm supp}/M_{\odot}).
\end{equation}

\section{The Luminosity Distribution Function}\label{sec:Ldist}

Our model results in an approximately log-normal distribution for the UV galaxy luminosities (1500 \AA) produced by halos of a given mass: 
\begin{equation}\label{eq:LDF}
\frac{dP}{dL}=\frac{1}{\sqrt{2\,\pi\,\sigma_{\rm L}^2}}\,{\rm exp}\left(-\frac{{\rm log}(L/L_{\rm c})}{2\,\sigma_{\rm L}^2}\right),
\end{equation}
in agreement with previous assumptions \citep[e.g.][]{CM05a, CM05b, CO06, VO04, VO06, VO08}.  As anticipated by the self-similarity of halo mergers \citep{Fakhouri10}, we find that, independent of redshift, $L_{\rm c}$ is proportional to halo mass.  \citet{Bouwens08a} previously estimated a power-law slope of 1.24 based on \citet{Bouwens07} data at $z \sim 4$.  We reiterate, that many of the assumptions that went into our model, including the lack of a quiescent component to star formation and dust extinction, are only valid at redshifts beyond 6.  We do not find a change in the proportionality of luminosity to halo mass at high masses as considered by \citet{Bouwens08a}.  Since the timescale for the coalescence of subhalos after merger is related to the ratio of their masses \citep{Wetzel09, WW10} and we have selected major merges based on a fixed mass ratio, we do not expect a fall off in the rate of major mergers in more massive halos.  Finally, we also find $\sigma_{\rm L}$ between 0.2 and 0.3 for all masses and redshifts considered.  Consequently, the galactic luminosity produced by halos of the same mass can easily vary by $\sim 1.5$ magnitudes or more.  For convenience, we will assume a fixed value of $\sigma_{\rm L}=0.25$ for all calculations of the LF throughout the rest of this paper.

\section{The Luminous Duty Cycle of Halos}\label{sec:eDC}

We first clarify a slight ambiguity of definition in the literature.  The duty cycle, $\epsilon_{\rm DC}$, is defined as the fraction of a halo's lifetime over which it is luminous.  If halos fluctuate stochastically on and off, then $\epsilon_{\rm DC}$ also represents the probability that a halo will be on at the moment of observation and the fraction of all halos at that time that are on.  Considering halos of a given mass for whom a single luminosity has been assumed, this concept of the duty cycle lowers the abundance of halos observed from that predicted in the halo mass function, i.e.
\begin{equation}\label{eq:eDC}
\frac{dn_{\rm obs}(L)}{dL}=\epsilon_{\rm DC}\,\frac{dn}{dM_{\rm h}}\,\frac{dL(M_{\rm h})}{dM_{\rm h}}.
\end{equation}
In various models, $\epsilon_{\rm DC}$ may be a function of variables such as mass or redshift or left as a constant parameter. 

\begin{figure}
\begin{center}
\includegraphics[width=\columnwidth]{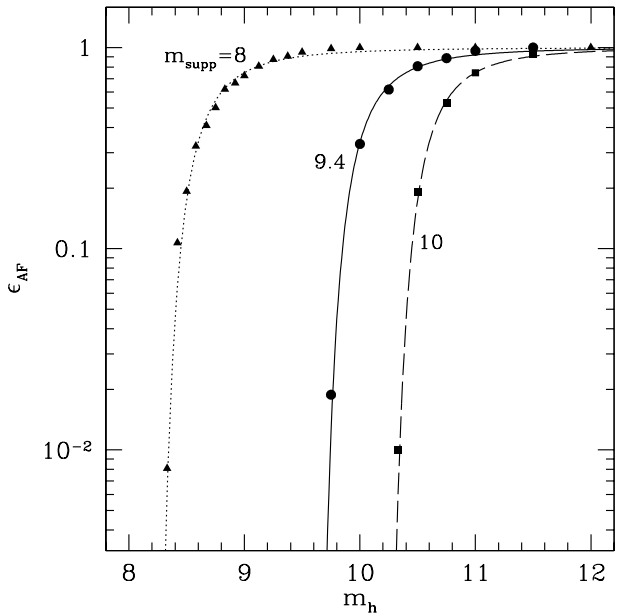}
\caption{\label{fig:eAF} 
The active fraction, $\epsilon_{\rm AF}$, or the fraction of halos that have had at least one starburst-generating merger in their lifetimes as a function of halo mass, $m_{\rm h}$, for three different values of the galaxy formation suppression threshold mass $M_{\rm supp}$.  Squares, circles, and triangles show merger-tree simulation results for $m_{\rm supp}=8$, 9.4, and 10, respectively, while dotted, solid, and long-dashed lines show the results from equation (\ref{eq:eAF}) for the same suppression mass values.  Enough merger histories we generated so that each point represents at least 100 active galaxies.
}
\end{center}
\end{figure}

However, if the luminosity of a halo or the frequency of its being in a luminous state is not constant in time, the fraction of observable halos for a given halo mass need not be equal to $\epsilon_{\rm DC}$.  In our model, there are two reasons why a halo may not be observable.  The first is that, given the continuous distribution in luminosity for a given halo mass, some are not bright enough to be detectable at their distance.  However, these halos will simply appear in another luminosity bin, and we will proceed by first calculating the full LF and subsequently applying an observable threshold for a given survey.  The second reason why a halo may not be observable is because, in the limited history of the universe at the moment of observation, its merger tree does not contain a single major merger whose constituents were more massive than $M_{\rm supp}$.  Thus, according to our model, it will not have had even one starburst, and will remain completely dark.  Halos much more massive than $M_{\rm supp}$ have had at least one starburst-generating merger, while those closer in mass to $M_{\rm supp}$ may not have since many of its recent progenitors are below $M_{\rm supp}$.  We define the probability that a halo of a given mass has had at least one starburst-generating merger as the ``active fraction," $\epsilon_{\rm AF}$. 

Using our merger tree code, we find a relation for the active fraction of halos as a function of mass that is nearly independent of redshift.  Since $\epsilon_{\rm AF}$ depends only on the distribution of merger histories, it is also independent of $\epsilon$ and the other details of our star formation model in \S\ref{sec:bursts}.  We show, in Figure ~\ref{fig:eAF}, $\epsilon_{\rm AF}$ as a function of mass calculated from our code for several values of $m_{\rm supp}$.  Each point was generated using enough random merger histories to produce at least 100 active galaxies.  A good fitting formula for $\epsilon_{\rm AF}$ given by:
\begin{equation}\label{eq:eAF}
{\rm log_{10}}\,\epsilon_{\rm AF}(m_{\rm h}, m_{\rm supp})=-\frac{1}{8\,(m_{\rm h}-m_{\rm supp})^{2.6}},
\end{equation}
where $m_{\rm h}$ and $m_{\rm supp}$ are defined by equation (\ref{eq:msupp}).  Throughout the rest of this paper, we rely on equation (\ref{eq:eAF}) to compute $\epsilon_{\rm AF}$ for continuous ranges of $m_{\rm h}$ and $m_{\rm supp}$.  Our results show that the abundance of halos hosting galaxies is suppressed even for halo masses up to an order-of-magnitude larger than the suppression threshold.  As we will see in \S\ref{sec:LF}, the large range of suppression masses combines with the distribution of possible luminosities for each halo to result in a gentle cut-off at the faint end of the LF rather than a sudden drop at a critical luminosity.  However, our model naturally reproduces a high value of $\epsilon_{\rm AF}$ for the largest halo masses, consistent with the rapid evolution of the halo mass function at $z\gtrsim6$ \citep{Trenti10}.

\begin{figure}
\begin{center}
\includegraphics[width=\columnwidth]{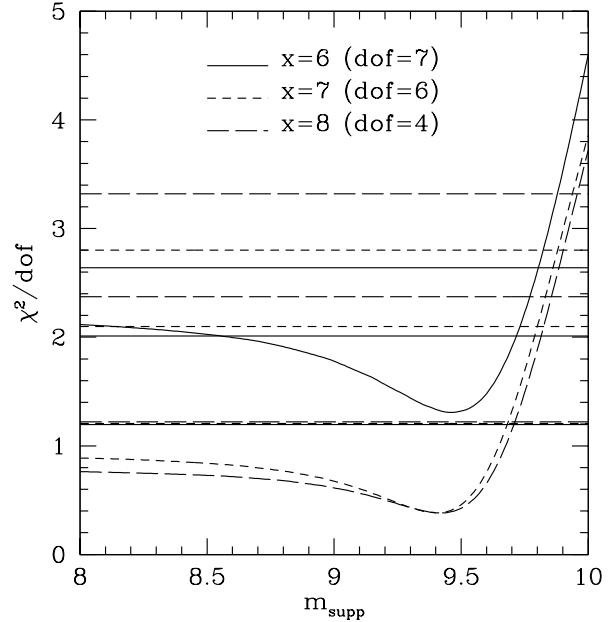}
\caption{\label{fig:chi2LF} 
The minimum reduced chi-squared (i.e. chi-squared per degree-of-freedom) marginalized over $L_{10}$ as a function of $m_{\rm supp}$.  Solid, short-dashed, and long-dashed lines show fits to the data at $z=6$, 7, and 8, respectively.  The bottom, middle, and top sets of horizontal lines denote the minimum reduced chi-squared values required for rejection with 70\%, 95\%, and 99\% confidence, respectively, for the number of degrees-of-freedom corresponding to the data at each redshift.
}
\end{center}
\end{figure}

\section{Fitting the Luminosity Function}\label{sec:LF}

We fit our model LF to the latest data available at $z=6-8$ from \citet{Bouwens07} and \citet{Bouwens10e} adopting the same magnitude conventions and ignoring, for simplicity, any bright-end upper limits.  All magnitudes we reference in this paper are rest-frame UV absolute magnitudes in the AB system.  We have calculated LFs at single redshifts for comparison with observations, ignoring for the time being the mass-dependent distribution of galaxies over the photometric redshift range of high-redshift surveys and its effects on the LF \citep{ML08b}.  At each redshift, we allow two fit parameters: $L_{10}=L_{\rm c}(M_{\rm h}=10^{10}\,\msun)$ and $M_{\rm supp}$.  $L_{10}$ is directly related to the star formation efficiency $\epsilon$ in our model for fixed choices of metallicity and IMF.  In Figure \ref{fig:chi2LF} we plot the minimum reduced chi-squared, $\chi^2_{\rm red}$, (i.e. chi-squared per degree of freedom) values matching the observed LF at $z=6$, 7, and 8 as a function of $M_{\rm supp}$.  Values of $L_{10}$ have been calculated to minimize $\chi^2_{\rm red}$ for each value of $M_{\rm supp}$.  

\begin{figure}
\begin{center}
\includegraphics[width=\columnwidth]{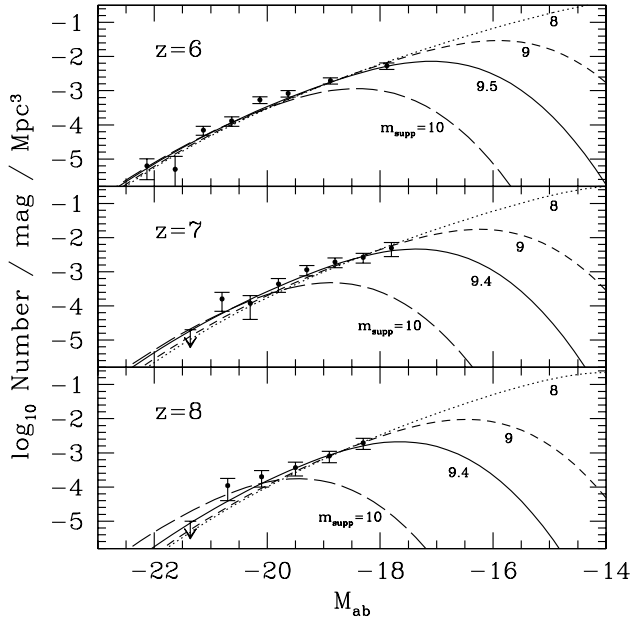}
\caption{\label{fig:LF} 
A comparison of our best-fit LFs to the data.  The top, middle, and bottom panels display results for $z=6$, 7, and 8, respectively.  The points and error-bars mark observations from \citet{Bouwens07, Bouwens10e}.  Dotted, short-dashed, and long-dashed curves are LFs assuming $m_{\rm supp}=8$, 9, and 10, respectively, with the best-fit value of $L_{10}$ for each value of $m_{\rm supp}$.  Finally, solid lines show results with the absolute minimum value of chi-square at each redshift.  The best-fit values of $m_{\rm supp}$ are 9.47, 9.4, and 9.42, for $z=6$, 7, and 8, respectively.
}
\end{center}
\end{figure}

A minimum in $\chi^2_{\rm red}$ appears at $m_{\rm supp} \approx 9.5$ at $z=6$, $\approx 9.4$ at $z=7$, and $\approx 9.42$ at $z=8$.  At $z=6$, all combinations of $L_{10}$ and $M_{\rm supp}$ are ruled out at the 70\% level.  However, values of $m_{\rm supp} < 8.55$ and $> 9.7$ are ruled out at the 95\% level, while $m_{\rm supp} > 9.8$ is ruled out with 99\% confidence.  At $z=7$ and $z=8$, no constraints are placed on the minimum value of $m_{\rm supp}$ at the 70\% level or stronger.  However, $m_{\rm supp} > 9.7$ (9.7), $> 9.8$ (9.85), and $> 9.9$ (9.95) are ruled out with 70\%, 95\%, and 99\% confidence, respectively, at $z=7$ (8).

These results clearly indicate that, while the masses of halos hosting observed LBGs are typically thought to be $> 10^{10}\,\msun$, lower luminosity galaxies must exist in halos smaller than $10^{10}\,\msun$, corresponding to a virial temperature of about $2\times10^{5}\,{\rm K}$, and very likely in ones at least as small as $5\times10^{9}\,\msun$ ($10^{5}\,{\rm K}$).  They also tentatively suggest that the minimum mass halo capable of hosting galaxies may be around $2.5\times10^{9}\,\msun$ ($7\times10^{4}\,{\rm K}$) with halos less massive than about $3.5\times10^{8}\,\msun$ ($1.7\times10^{4}\,{\rm K}$) unable to host galaxies with some confidence given the data at $z=6$.

Chi-squared is minimized when $L_{10} \approx 27.2$, 27.4, and 27.7 at $z=6$, 7, and 8, respectively.  For our choices of metallicity and IMF, these values imply that galaxy formation is relatively inefficient, with very small fractions of galactic gas (0.2\%, 0.4\%, and 0.5\% for each redshift) being converted into stars per dynamical time.

Our best-fit LFs, along with ones assuming $m_{\rm supp}=8$, 9, and 10, are shown for each redshift in Figure \ref{fig:LF}.  The data from \citet{Bouwens07, Bouwens10e} are plotted for comparison.  The best-fit model deviates qualitatively from a Schechter function fit outside the observed magnitude range.  At the bright end, for magnitudes $< -21$, our predicted LF remains much flatter than a Schechter fit, which is already beginning to drop exponentially.  The shallower slope is due to two effects: first, the exponential tail of the halo mass function falls off more slowly with increasing mass than a Schechter function with luminosity proportional to mass, and second, the large spread in the luminosity permitted for each halo mass allows abundant, smaller halos emitting at higher than average luminosity to bolster the population of bright galaxies.  

On the other hand, the suppression of star formation drastically reduces the abundance of galaxies at magnitudes fainter than currently observable compared with expectations from a simple extrapolation of the Schechter function.  The result is a flatter LF slope in the observed region for increasing $m_{\rm supp}$.  Figure \ref{fig:LF} clearly illustrates the disparate predictions for the abundance of faint galaxies between different fiducial values of $M_{\rm supp}$.  Additional data at only about a magnitude fainter than the current observational threshold will greatly improve our ability to constrain the minimum halo mass capable of forming galaxies.  For reference, observations down to a magnitude of -16.8 at $z=7$ will require a sensitivity of about $1.5\,{\rm nJy}$, close to what is expected with {\it {JWST}}.  However, while the $1\sigma$ errors in the current data were calculated based on the shot noise and cosmic variance from an amalgam of observations from several different fields, we conservatively estimate the $1\sigma$ error on the abundance at this magnitude in a single $2'\times2'$ pointing of NIRCam on {\it{JWST}} to be about 50\% \citep{Munoz10}.

We have explicitly ignored the influence of quiescent star formation in our model.  Such a mechanism in small halos may reduce the effects that we describe of a galaxy suppression mass on the faint end of the LF making it more difficult to probe such physics with future surveys.  However, work by \citet{Lacey10} has shown that merger-driven starbursts do dominate the UV LF down to at least magnitude -17 at z=6 and to -15 by z=10, albeit with a very different IMF.  Thus, while more complicated models may be required for high-precision measurements of $M_{\rm supp}$ even with a complete LF, we are confident that deeper surveys of the not-to-distant future will help illuminate some of the physics of galaxy suppression.

\section{Star Formation Rate}\label{sec:sfr}

The SFR of high-redshift galaxies is important for understanding the star formation history of the Universe \citep{Madau98} and the ionization state of the IGM \citep{Madau99}.  Its estimation relies on a proportionality between UV luminosity and SFR based on two assumptions: an exponential burst of star formation has a timescale, $\tau$, that is longer than 1 Gyr, and the stellar population is observed after one exponential time scale has past \citep{Madau98}.  However, if the age of the universe is shorter than 1 Gyr, at least one of these assumptions must be violated. 

For the best-fit star formation efficiencies we found from the data at $z=6$, 7, and 8, the typical exponential starburst timescale given by equation (\ref{eq:tau7}) is of order $\tau \sim 10\,{\rm Gyr}$, an order-of-magnitude or more longer than the age of the universe.  Equation (\ref{eq:Msfr7}) gives the typical SFR to be of order $1\,{\rm \msun/yr}$ in a burst with $10^{10}\,\msun$ worth of gas remaining; if the amount of initial gas in a halo as a fraction of the total halo mass is $\Omega_{\rm b}/\Omega_{\rm m} \approx 0.16$, this corresponds to the initial SFR in a halo of about $6.25\times10^{10}\,{\rm \msun}$.

\begin{figure}
\begin{center}
\includegraphics[width=\columnwidth]{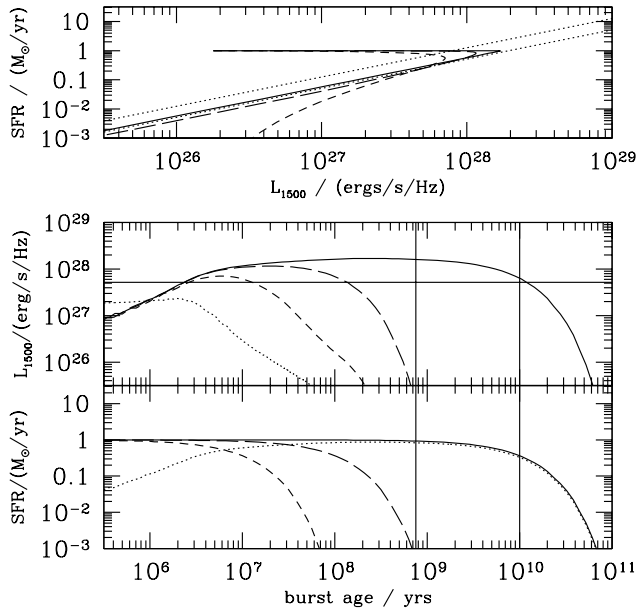}
\caption{\label{fig:burst} 
The UV luminosity and SFR evolution of exponential bursts with $(\tau/{\rm Gyr},\dot{M}_{\star}(t=0)\,{\rm \msun^{-1}\,yr})=(10,1)$, (0.1,3), and (0.01,10) denoted by solid, long-dashed, and short-dashed curves, respectively.  The top panel tracks the bursts in SFR-$L_{1500}$ space.  Here, the upper and lower dotted lines show a proportional relationship between SFR and $L_{1500}$ with constants of $8\times10^{27}\,{\rm erg/s/Hz/(\msun/yr)}$ and $2\times10^{28}\,{\rm erg/s/Hz/(\msun/yr)}$, respectively.  The middle panel shows the burst lightcurves with right and left vertical lines denoting the $\tau=10\,{\rm Gyr}$ and the age of the universe at $z=7$, respectively.  The lightcurve for an instantaneous burst producing $10^6\,\msun$ worth of stars is given by the dotted curve.  The horizontal line marks the observable threshold at a magnitude of -18.  The bottom panel plots the evolution of the SFR with time, while the dotted curve here shows the SFR expected from by the relationship between SFR and $L_{1500}$ given the burst luminosity as a function of time.
}
\end{center}
\end{figure}

Figure \ref{fig:burst} shows the evolution of the luminosity at 1500 \AA, $L_{1500}$, and SFR, $\dot{M}_{\star}$, with time and their relationship calculated for exponential bursts using a simple stellar population from Starburst99 \citep{Leitherer99}.  Solid lines represent typical bursts in high-redshift galaxies with the initial SFR set at $1\,{\rm \msun/yr}$ and the exponential time scale $\tau = 10\,{\rm Gyr}$.  For $t>\tau$, both $\dot{M}_{\star}$ and $L_{1500}$ decrease exponentially over time with timescale $\tau$ so that $L_{1500}$ is proportional to $\dot{M}_{\star}$.  This is because the exponential timescale is much longer than the lifetime of the stars that dominate the UV luminosity.  The amplitude of the relation is set by the IMF and metallicity of the stellar population; for the choices described in \S\ref{sec:bursts}, we find approximately $L_{1500}=2\times10^{28}\,(\dot{M}_{\star}/{\rm \msun\,yr})\,{\rm erg/s/Hz}$ with a proportionality constant a factor of 2.5 different than the $8\times10^{27}\,{\rm erg/s/Hz/(\msun/yr)}$.  However, before $t=\tau$, the luminosity is still rising with increasing time, while the SFR remains essentially unchanged.  Since the age of the universe is much less than $\tau$, all bursts are observed in this phase before the $L_{1500}-\dot{M}_{\star}$ proportionality has stabilized.  

\begin{figure}
\begin{center}
\includegraphics[width=\columnwidth]{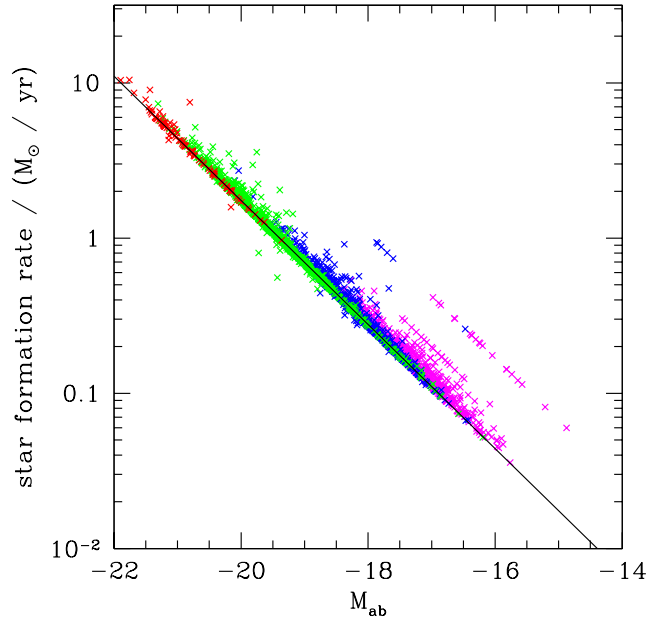}
\caption{\label{fig:sfrL7} 
The SFR vs. rest-frame UV magnitude of simulated halos.  Magenta, blue, green, and red points denote halos of mass $10^{10}$, $10^{10.5}$, $10^{11}$, and $10^{11.5}\,\msun$, respectively.  The solid, black line marks $L_{1500}=2\times10^{28}\,(\dot{M}_{\star}/{\rm \msun\,yr})\,{\rm erg/s/Hz}$.
}
\end{center}
\end{figure}

Thus, the SFR will typically be somewhat higher than that inferred from the $L_{1500}-\dot{M}_{\star}$ proportionality.  The ratio between the true and expected SFRs will depend on how close the burst is to its maximum luminosity, the point where the expected SFR is approximately equal to its initial value.  The burst lightcurve is relatively flat near its maximum value over the time approximately $10^7-10^9\,{\rm yrs}$ after it begins.  If the burst is more than $10^7\,{\rm yrs}$ old at the time of observation, the difference between the true and expected SFRs will not be very significant.  A burst observed at $z=7$ is $10^7\,{\rm yrs}$ old if it started at $z \approx 7.07$.

For completion, we also show in Figure \ref{fig:burst} the evolution of $L_{1500}$ and $\dot{M}_{\star}$ for bursts with $\tau=0.1$ and 0.01 Gyr, less than the 1 Gyr minimum considered by \citet{Madau98}.  These timescales are achieved at $z=7$ if for combinations of the star formation efficiency and the disk spin parameter such than $\epsilon^{-1}\,\lambda=0.17$ and 0.017, respectively.  The luminosity in each case begins to decline before reaching the maximum it would have achieved had $\tau$ been longer.  The fall-off in luminosity is only slightly slower than exponential for $\tau=0.1\,{\rm Gyr}$ so that the SFR and UV luminosity reach a nearly proportional relationship after $t=\tau$, albeit with a coefficient slightly higher than that seen for higher $\tau$.  However, the decline in luminosity is more power-law than exponential for $\tau=0.01\,{\rm Gyr}$ leading to a very non-linear relationship between SFR and luminosity after $t=\tau$.  In both cases, the SFR is much less than expected for a given luminosity.  This is because the timescale $\tau$ is not so much longer than the lifetimes of the stars that dominate the UV luminosity.  Luminosity from stars produced at $t=\tau$, for example, contribute significantly to the luminosity at $0.1\,\tau$, whereas the contribution would be completely negligible for much larger $\tau$.  Consequently, the luminosity for a given instantaneous SFR can be much higher than expected.

Using our merger tree and star formation code, we calculate the instantaneous $\dot{M}_{\star}$ at the time of observation for each of our modeled galaxies and test the accuracy of the \citet{Madau98} proportionality at $z = 7$ over a wide range of halo masses and for a full distribution of spin parameters and merger histories.  Figure \ref{fig:sfrL7} shows the relationship between instantaneous SFR, $\dot{M}_{\star}$, and UV luminosity, $L_{1500}$, for galaxies in halos at $z=7$ with masses of $10^{10}$, $10^{10.5}$, $10^{11}$, and $10^{11.5}\,\msun$.  We have set $m_{\rm supp} = 9.4$.  Each point represents a single halo, and where more than one ongoing starburst is present, we have simply added the contributing SFRs.  The discrete "lines" of points above the main body for the smaller halo masses is a resolution effect of our code; while their exact positions should not be taken as precise, such points do represent a population of halos lying above the standard $L_{1500}-\dot{M}_{\star}$ relation.

The majority of points in Figure \ref{fig:sfrL7} do show a rough proportionality between $L_{1500}$ and $\dot{M}_{\star}$.  However, the proportionality constant is slightly higher than the $2\times10^{28}\,{\rm erg/s/Hz/(\msun/yr)}$ value for bursts with ages longer than their exponential time scale with the difference depending on halo mass.  Lower mass halos tend to be populated by younger bursts that are further from reaching their maximum luminosity than higher mass halos.  If we constrain $L_{1500} \propto \dot{M}_{\star}$, we find a proportionality constant of $1.7\times10^{28}\,{\rm erg/s/Hz/(\msun/yr)}$ for $10^{10}\,\msun$ halos which estimates SFRs to be about 15\% higher than the constant for older bursts.  Given the typical uncertainties in measuring the total star formation rate at high redshift -- sample completeness, cosmic variance, uncertain IMF and metallicity, etc. -- an additional $\sim 20\%$ error will not significantly affect current estimates.

\section{Ionization State of the IGM}\label{sec:IGM}

After cosmic reionization, the ionization state of the IGM depends on the balance between the recombination rate and the production rate of ionizing photons.  On its own, the formation of stars in galaxies can maintain the ionization of the IGM through its production of ionizing photons if the star formation rate density (SFRD) is higher than a critical value given by \citet{Madau99}:
\begin{equation}\label{eq:sfrd}
\dot{\rho}_{\star} \approx 2\times10^{-3}\,f_{\rm esc}^{-1}\,C\,\left(\frac{1+z}{10}\right)^3\,{\rm \msun/yr/Mpc^3},
\end{equation}
where $f_{\rm esc}$ is the fraction of ionizing photons produced in galaxies that escape into the IGM, and $C$ is the IGM clumping factor. 

\begin{figure}
\begin{center}
\includegraphics[width=\columnwidth]{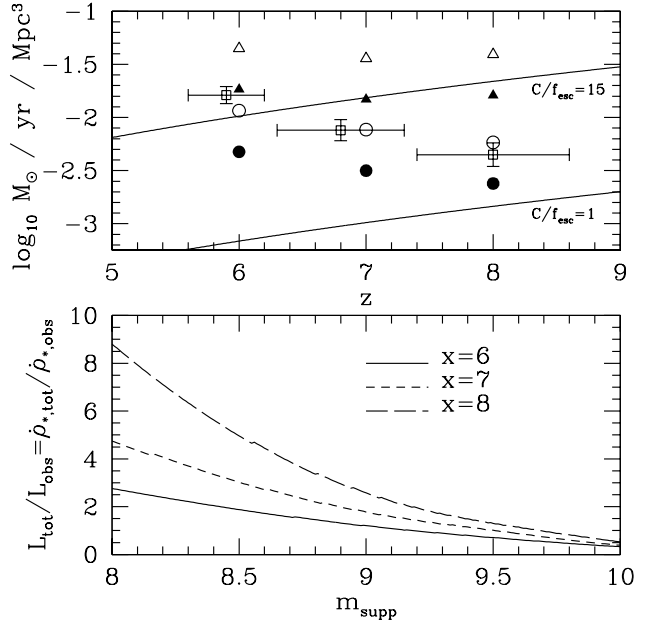}
\caption{\label{fig:sfrd} 
The top panel shows the SFRD produced by the total galaxy population at $z=6$, 7, and 8.  Circles denote results using best-fit values of $M_{\rm supp}$ and $L_{10}$ at each redshift, while triangles assume $M_{\rm supp}=10^8\,\msun$ and the corresponding best-fit values of $L_{10}$.  Filled (empty) points use $L_{1500}/\dot{M}_{\star}=2\times10^{28}\,(8\times10^{27})\,{\rm erg/s/Hz/(\msun/yr)}$.  Square points with error bars denote observed values from \citet{Bouwens07} and \citet{Bouwens10e}.  The minimum SFRD required to keep the IGM ionized as given by Eq. (\ref{eq:sfrd}) for $f_{\rm esc}^{-1}\,C=15$ and 1 are shown by the upper and lower solid lines, respectively.  The bottom panel shows the ratio of the total UV luminosity or SFRD to the \citet{Bouwens07} and \citet{Bouwens10e} observations as a function of $M_{\rm supp}$.  The solid, short-dashed, and long-dashed lines denote $z=6$, 7, and 8, respectively.
}
\end{center}
\end{figure}

Using the standard $L_{1500}$ to $\dot{M}_{\star}$ conversion, recent observational studies have estimated the currently observable SFRD to be just enough to keep the universe ionized if $f_{\rm esc}^{-1}\,C=1$.  However, much of the star formation below the observable threshold is not included.  In Figure \ref{fig:sfrd}, we compare our calculations for the total SFRD at $z=6$, 7, and 8 assuming the best-fit values of $L_{10}$ and $M_{\rm supp}$ at each redshift to the observed estimates and to equation (\ref{eq:sfrd}).  We show results for $L_{1500}$ to $\dot{M}_{\star}$ ratios of both $8\times10^{27}\,{\rm erg/s/Hz/(\msun/yr)}$ (the typically used value) and $2\times10^{28}\,{\rm erg/s/Hz/(\msun/yr)}$ (consistent with our choices of IMF and metallicity).  We also show the factor by which the total values of the SFRD and UV luminosity exceed those observed by \citet{Bouwens07} and \citet{Bouwens10e}.  Factors less than unity indicate that the observed points are higher than average in the universe at that redshift due to Poisson fluctuations or cosmic variance so that the observed SFRD is higher than the average over the whole population.

Our results show that the ability of the total galaxy population to account for the UV background required to keep the IGM ionized depends critically on the value of $m_{\rm supp}$.  For $m_{\rm supp}=8$, the total SFRD or UV luminosity is about 3--9 times the observed values with more star formation and luminosity missing at higher redshift.  However, at the best-fit values of $m_{\rm supp}=9.5$ at $z=6$ and 9.4 at $z=7$ and 8, the galaxy population produces no more SFRD than observed (and somewhat less for $z=6$).   Assuming $f_{\rm esc}^{-1}\,C=1$, the total SFRD for all parameters and redshifts considered meet the requirement for maintaining the ionization of the IGM.  However, if $f_{\rm esc}^{-1}\,C=15$ (e.g. $f_{\rm esc}=0.2$ and $C=3$), the galaxy population can keep the IGM at $z=8$ ionized only for a choice of IMF and metallicity that gives the \citet{Madau98} ratio of $L_{1500}$ to $\dot{M}_{\star}$ ratio of $3\times10^{27}\,{\rm erg/s/Hz/(\msun/yr)}$ and if $m_{\rm supp}\sim8$, lower than our best-fit value.  Finally, since the amount of star formation below the observable limit increases with redshift, we find that the evolution of the total SFRD with redshift is much flatter than that observed.

\section{Conclusions}\label{sec:conc}

In this paper, we combine a standard merger tree algorithm with a simple star formation prescription designed to encapsulate the main physical processes relevant at $z\gtrsim6$.  Our model both accounts for a range of possible galaxy luminosities for each halo mass and includes a sharp galaxy formation cut-off in halo mass below $M_{\rm supp}$.

\begin{itemize}
\item {
We confirm that the luminosity distribution function for halos of a given mass is a roughly log-normal distribution with a variance of $\sim 1.5$ magnitudes and a proportional relationship between the mean luminosity and halo mass (see \S\ref{sec:Ldist}).
}
\end{itemize}

At a fixed halo mass of $10^{10}\,\msun$, the mean log-luminosities are ${\rm log_{10}}(L_{10}\,{\rm erg^{-1}\,s\,Hz})=27.2$, 27.4, and 27.7 at $z=6$, 7, and 8, respectively, suggesting that for a fixed halo mass, galaxies are brighter on average at higher redshift, consistent with results from Schechter fits.  However, while the exponential tail of the high-redshift halo mass function is less sharp than that of a Schechter function, the range of possible luminosities for a fixed halo mass further slows the fall-off of the predicted galaxy LF at the bright end.  While still being consistent with the data, our shallower LF anticipates the discovery of a larger population of very bright galaxies at $z=7$ and 8 as survey fields increase in size. 

\begin{itemize}
\item {
We also show that an active fraction of halos that approaches unity with increasing halo mass can be naturally explained by a suppression halo mass for galaxy formation combined with the variety of possible merger histories (see \S\ref{sec:eDC}).
}
\end{itemize}

This active fraction is well-approximated by the formula given in Eq. (\ref{eq:eAF}).  One can easily use this expression, along with our log-normal distributions of UV luminosity for each halo mass to calculate the galaxy LF from the halo mass function.  The resulting LF does not have a sharp cutoff at the faint end but rather turns over gently.  Thus, we predict that as long as future observations show a LF that increase with ever decreasing luminosity, the surveyed region will never be volume-complete.

\begin{itemize}
\item {
The current data suggests that the minimum mass halo capable of hosting galaxies may be around $2.5\times10^{9}\,\msun$, corresponding to a virial temperature of $7\times10^{4}\,{\rm K}$ (see \S\ref{sec:LF}).
}
\end{itemize}

We find a strong upper limit of $M_{\rm supp}<6\times10^{9}\,\msun$ ($10^{5}\,{\rm K}$) with at least 95\% confidence.  However, lower limits from the current data are quite weak with halos less massive than about $3.5\times10^{8}\,\msun$ ($1.7\times10^{4}\,{\rm K}$) unable to host galaxies with some confidence given the data at $z=6$.

\begin{itemize}
\item {
We find a best-fit star formation efficiency at high redshift of approximately 0.2-0.5\% per dynamical time, implying a starburst exponential time scale much longer than the age of the universe.
}
\end{itemize}

A more top-heavy IMF would have required even less efficient star formation, corresponding to even longer burst time scales, to produce the same observed luminosities.  However, the long burst time scale does not create lightbulb-like galaxies that, once switched on, are always emitting with the same luminosity.  Instead continued merger activity disrupts old bursts and replaces them with new ones based on the particular history of the host halo.

\begin{itemize}
\item {
We show that the proportionality of $L_{1500}$ to $\dot{M}_{\star}$ is usually an adequate approximation (see \S\ref{sec:sfr}).
}
\end{itemize}

While the \citet{Madau98} proportionality relies on long-lived bursts in the tail of their exponentially decreasing rate of star formation, most bursts at $z=7$ are emitting near their maximum luminosity where the track of SFR vs. UV luminosity begins to join the proportional relationship.  Despite their young ages compared to their exponential time scale, this is because the bursts are typically older than $10^7\,{\rm yrs}$ at the time of observation, old enough that the massive stars providing the bulk of the UV luminosity are beginning to die out as fast as new ones are added.  Although the lowest mass halos may host very young bursts that have somewhat higher SFRs than expected for their luminosities, using a standard proportionality of $L_{1500}$ to $\dot{M}_{\star}$ adds additional errors of only tens of percent.  However, some care must be taken 
in selecting a constant of proportionality consistent with specific choices of metallicity and IMF rather than using the \citet{Madau98} value indiscriminately.  Additionally, since bursts are likely to remain close to their maximum SFRs and luminosities for most of their lifetimes, ongoing accretion between major mergers is less likely to be important.

\begin{itemize}
\item {
When extrapolated down to faint luminosities below the current observable threshold, the total SFRD of the galaxy population will only at most 3--9 times higher than what has already been observed even if the minimum halo mass forming galaxies is as low as $10^8\,\msun$ (see \S\ref{sec:IGM}).
}
\end{itemize}

The gentle turnover at the faint end of the LF, even given a sharp cutoff in the halo mass capable of producing galaxies, results in less star formation below the observable limit than if the LF dropped sharply at the mean luminosity corresponding to the same halo mass.  While the total galaxy population with $m_{\rm supp}=8$ may be able to keep the IGM ionized given $f_{\rm esc}^{-1}\,C \sim 15$, for our best-fit value of $m_{\rm supp}\approx9.4$, no significant star formation lies below a rest-frame UV magnitude of -18.  In such a case, galaxies may only be responsible for maintaining the ionization state of the IGM if $f_{\rm esc}^{-1}\,C \sim 1$.  Interestingly, we also find that, since the amount of missing star formation increases with redshift, the redshift evolution of the total star formation history of the universe is flatter than observed.
 
Although the current data from LBG drop-outs does not place a strong
lower-limit on the minimum halo mass required to host galaxies at
redshifts $\gtrsim 6$, we have shown that {\it{JWST}} and other future
deep surveys will provide much tighter constraints.  These results
will not only shed light on the contribution of galaxies to the UV
background that keeps the IGM ionized but also hint at the feedback
physics that limits galaxy formation.

\section{Acknowledgements}

We thank Steve Furlanetto for useful discussions.  This work was supported in part by NSF grant AST-0907890 and NASA grants NNX08AL43G and NNA09DB30A (for A.L.).



\begin{thebibliography}{45}
\expandafter\ifx\csname natexlab\endcsname\relax\def\natexlab#1{#1}\fi

\bibitem[{{Barkana} \& {Loeb}(2000{\natexlab{a}})}]{BL00a}
{Barkana}, R., \& {Loeb}, A. 2000{\natexlab{a}}, \apj, 531, 613

\bibitem[{{Barkana} \& {Loeb}(2000{\natexlab{b}})}]{BL00b}
---. 2000{\natexlab{b}}, \apj, 539, 20

\bibitem[{{Baugh} {et~al.}(2005){Baugh}, {Lacey}, {Frenk}, {Granato}, {Silva},
  {Bressan}, {Benson}, \& {Cole}}]{Baugh05}
{Baugh}, C.~M., {Lacey}, C.~G., {Frenk}, C.~S., {Granato}, G.~L., {Silva}, L.,
  {Bressan}, A., {Benson}, A.~J., \& {Cole}, S. 2005, \mnras, 356, 1191

\bibitem[{{Bouwens} {et~al.}(2007){Bouwens}, {Illingworth}, {Franx}, \&
  {Ford}}]{Bouwens07}
{Bouwens}, R.~J., {Illingworth}, G.~D., {Franx}, M., \& {Ford}, H. 2007, \apj,
  670, 928

\bibitem[{{Bouwens} {et~al.}(2008)}]{Bouwens08a}
{Bouwens}, R.~J., {et~al.} 2008, \apj, 686, 230

\bibitem[{{Bouwens} {et~al.}(2010{\natexlab{a}}){Bouwens}, {Illingworth},
  {Oesch}, {Trenti}, {Stiavelli}, {Carollo}, {Franx}, {van Dokkum},
  {Labb{\'e}}, \& {Magee}}]{Bouwens10a}
{Bouwens}, R.~J., {Illingworth}, G.~D., {Oesch}, P.~A., {Trenti}, M.,
  {Stiavelli}, M., {Carollo}, C.~M., {Franx}, M., {van Dokkum}, P.~G.,
  {Labb{\'e}}, I., \& {Magee}, D. 2010{\natexlab{a}}, \apjl, 708, L69

\bibitem[{{Bouwens} {et~al.}(2010{\natexlab{b}}){Bouwens}, {Illingworth}, {Labbe}, {Oesch},
  {Carollo}, {Trenti}, {van Dokkum}, {Franx}, {Stiavelli}, {Gonzalez}, \&
  {Magee}}]{Bouwens10c}
{Bouwens}, R.~J., {Illingworth}, G.~D., {Labbe}, I., {Oesch}, P.~A., {Carollo},
  M., {Trenti}, M., {van Dokkum}, P.~G., {Franx}, M., {Stiavelli}, M.,
  {Gonzalez}, V., \& {Magee}, D. 2010{\natexlab{b}}, ArXiv e-prints

\bibitem[{{Bouwens} {et~al.}(2010{\natexlab{c}}){Bouwens}, {Illingworth},
  {Oesch}, {Labbe}, {Trenti}, {van Dokkum}, {Franx}, {Stiavelli}, {Carollo},
  {Magee}, \& {Gonzalez}}]{Bouwens10e}
{Bouwens}, R.~J., {Illingworth}, G.~D., {Oesch}, P.~A., {Labbe}, I., {Trenti},
  M., {van Dokkum}, P., {Franx}, M., {Stiavelli}, M., {Carollo}, C.~M.,
  {Magee}, D., \& {Gonzalez}, V. 2010{\natexlab{c}}, ArXiv e-prints

\bibitem[{{Cole} {et~al.}(2000){Cole}, {Lacey}, {Baugh}, \& {Frenk}}]{Cole00}
{Cole}, S., {Lacey}, C.~G., {Baugh}, C.~M., \& {Frenk}, C.~S. 2000, \mnras,
  319, 168

\bibitem[{{Cooray} \& {Milosavljevi{\'c}}(2005{\natexlab{a}})}]{CM05a}
{Cooray}, A., \& {Milosavljevi{\'c}}, M. 2005{\natexlab{a}}, \apjl, 627, L85

\bibitem[{{Cooray} \& {Milosavljevi{\'c}}(2005{\natexlab{b}})}]{CM05b}
---. 2005{\natexlab{b}}, \apjl, 627, L89

\bibitem[{{Cooray} \& {Ouchi}(2006)}]{CO06}
{Cooray}, A., \& {Ouchi}, M. 2006, \mnras, 369, 1869

\bibitem[{{Dunkley} {et~al.}(2009){Dunkley}, {Komatsu}, {Nolta}, {Spergel},
  {Larson}, {Hinshaw}, {Page}, {Bennett}, {Gold}, {Jarosik}, {Weiland},
  {Halpern}, {Hill}, {Kogut}, {Limon}, {Meyer}, {Tucker}, {Wollack}, \&
  {Wright}}]{Dunkley09}
{Dunkley}, J., {Komatsu}, E., {Nolta}, M.~R., {Spergel}, D.~N., {Larson}, D.,
  {Hinshaw}, G., {Page}, L., {Bennett}, C.~L., {Gold}, B., {Jarosik}, N.,
  {Weiland}, J.~L., {Halpern}, M., {Hill}, R.~S., {Kogut}, A., {Limon}, M.,
  {Meyer}, S.~S., {Tucker}, G.~S., {Wollack}, E., \& {Wright}, E.~L. 2009,
  \apjs, 180, 306

\bibitem[{{Fakhouri} {et~al.}(2010){Fakhouri}, {Ma}, \&
  {Boylan-Kolchin}}]{Fakhouri10}
{Fakhouri}, O., {Ma}, C., \& {Boylan-Kolchin}, M. 2010, \mnras, 406, 2267

\bibitem[{{Finlator} {et~al.}(2010){Finlator}, {Oppenheimer}, \&
  {Dav{\'e}}}]{Finlator10}
{Finlator}, K., {Oppenheimer}, B.~D., \& {Dav{\'e}}, R. 2010, ArXiv e-prints

\bibitem[{{Gonz{\'a}lez} {et~al.}(2010){Gonz{\'a}lez}, {Labb{\'e}}, {Bouwens},
  {Illingworth}, {Franx}, {Kriek}, \& {Brammer}}]{Gonzalez10}
{Gonz{\'a}lez}, V., {Labb{\'e}}, I., {Bouwens}, R.~J., {Illingworth}, G.,
  {Franx}, M., {Kriek}, M., \& {Brammer}, G.~B. 2010, \apj, 713, 115

\bibitem[{{Kennicutt}(1998)}]{Kennicutt98}
{Kennicutt}, Jr., R.~C. 1998, \apj, 498, 541

\bibitem[{{Komatsu} {et~al.}(2009){Komatsu}, {Dunkley}, {Nolta}, {Bennett},
  {Gold}, {Hinshaw}, {Jarosik}, {Larson}, {Limon}, {Page}, {Spergel},
  {Halpern}, {Hill}, {Kogut}, {Meyer}, {Tucker}, {Weiland}, {Wollack}, \&
  {Wright}}]{Komatsu09}
{Komatsu}, E., {Dunkley}, J., {Nolta}, M.~R., {Bennett}, C.~L., {Gold}, B.,
  {Hinshaw}, G., {Jarosik}, N., {Larson}, D., {Limon}, M., {Page}, L.,
  {Spergel}, D.~N., {Halpern}, M., {Hill}, R.~S., {Kogut}, A., {Meyer}, S.~S.,
  {Tucker}, G.~S., {Weiland}, J.~L., {Wollack}, E., \& {Wright}, E.~L. 2009,
  \apjs, 180, 330

\bibitem[{{Labb{\'e}} {et~al.}(2010{\natexlab{a}}){Labb{\'e}}, {Gonz{\'a}lez},
  {Bouwens}, {Illingworth}, {Oesch}, {van Dokkum}, {Carollo}, {Franx},
  {Stiavelli}, {Trenti}, {Magee}, \& {Kriek}}]{Labbe10a}
{Labb{\'e}}, I., {Gonz{\'a}lez}, V., {Bouwens}, R.~J., {Illingworth}, G.~D.,
  {Oesch}, P.~A., {van Dokkum}, P.~G., {Carollo}, C.~M., {Franx}, M.,
  {Stiavelli}, M., {Trenti}, M., {Magee}, D., \& {Kriek}, M.
  2010{\natexlab{a}}, \apjl, 708, L26

\bibitem[{{Labb{\'e}} {et~al.}(2010{\natexlab{b}}){Labb{\'e}}, {Gonz{\'a}lez},
  {Bouwens}, {Illingworth}, {Franx}, {Trenti}, {Oesch}, {van Dokkum},
  {Stiavelli}, {Carollo}, {Kriek}, \& {Magee}}]{Labbe10b}
{Labb{\'e}}, I., {Gonz{\'a}lez}, V., {Bouwens}, R.~J., {Illingworth}, G.~D.,
  {Franx}, M., {Trenti}, M., {Oesch}, P.~A., {van Dokkum}, P.~G., {Stiavelli},
  M., {Carollo}, C.~M., {Kriek}, M., \& {Magee}, D. 2010{\natexlab{b}}, \apjl,
  716, L103

\bibitem[{{Lacey} {et~al.}(2010){Lacey}, {Baugh}, {Frenk}, {Benson}, \&
  {.}}]{Lacey10}
{Lacey}, C.~G., {Baugh}, C.~M., {Frenk}, C.~S., {Benson}, A.~J., \& {.} 2010,
  ArXiv e-prints

\bibitem[{{Leitherer} {et~al.}(1999){Leitherer}, {Schaerer}, {Goldader},
  {Gonz{\'a}lez Delgado}, {Robert}, {Kune}, {de Mello}, {Devost}, \&
  {Heckman}}]{Leitherer99}
{Leitherer}, C., {Schaerer}, D., {Goldader}, J.~D., {Gonz{\'a}lez Delgado},
  R.~M., {Robert}, C., {Kune}, D.~F., {de Mello}, D.~F., {Devost}, D., \&
  {Heckman}, T.~M. 1999, \apjs, 123, 3

\bibitem[{{Madau} {et~al.}(1999){Madau}, {Haardt}, \& {Rees}}]{Madau99}
{Madau}, P., {Haardt}, F., \& {Rees}, M.~J. 1999, \apj, 514, 648

\bibitem[{{Madau} {et~al.}(1998){Madau}, {Pozzetti}, \& {Dickinson}}]{Madau98}
{Madau}, P., {Pozzetti}, L., \& {Dickinson}, M. 1998, \apj, 498, 106

\bibitem[{{Mo} {et~al.}(1998){Mo}, {Mao}, \& {White}}]{MMW98}
{Mo}, H.~J., {Mao}, S., \& {White}, S.~D.~M. 1998, \mnras, 295, 319

\bibitem[{{Mu{\~n}oz} \& {Loeb}(2008{\natexlab{a}})}]{ML08a}
{Mu{\~n}oz}, J.~A., \& {Loeb}, A. 2008{\natexlab{a}}, \mnras, 385, 2175

\bibitem[{{Mu{\~n}oz} \& {Loeb}(2008{\natexlab{b}})}]{ML08b}
---. 2008{\natexlab{b}}, \mnras, 386, 2323

\bibitem[{{Mu{\~n}oz} {et~al.}(2010){Mu{\~n}oz}, {Trac}, \& {Loeb}}]{Munoz10}
{Mu{\~n}oz}, J.~A., {Trac}, H., \& {Loeb}, A. 2010, \mnras, 405, 2001

\bibitem[{{Oesch} {et~al.}(2010){Oesch}, {Bouwens}, {Carollo}, {Illingworth},
  {Trenti}, {Stiavelli}, {Magee}, {Labb{\'e}}, \& {Franx}}]{Oesch10b}
{Oesch}, P.~A., {Bouwens}, R.~J., {Carollo}, C.~M., {Illingworth}, G.~D.,
  {Trenti}, M., {Stiavelli}, M., {Magee}, D., {Labb{\'e}}, I., \& {Franx}, M.
  2010, \apjl, 709, L21

\bibitem[{{Overzier} {et~al.}(2009){Overzier}, {Guo}, {Kauffmann}, {De Lucia},
  {Bouwens}, \& {Lemson}}]{Overzier09}
{Overzier}, R.~A., {Guo}, Q., {Kauffmann}, G., {De Lucia}, G., {Bouwens}, R.,
  \& {Lemson}, G. 2009, \mnras, 394, 577

\bibitem[{{Rai{\v c}evi{\'c}} {et~al.}(2010){Rai{\v c}evi{\'c}}, {Theuns}, \&
  {Lacey}}]{Raicevic10}
{Rai{\v c}evi{\'c}}, M., {Theuns}, T., \& {Lacey}, C. 2010, ArXiv e-prints

\bibitem[{{Robertson}(2010{\natexlab{a}})}]{Robertson10a}
{Robertson}, B.~E. 2010{\natexlab{a}}, \apjl, 713, 1266

\bibitem[{{Robertson}(2010{\natexlab{b}})}]{Robertson10b}
---. 2010{\natexlab{b}}, \apj, 716, L229

\bibitem[{{Salvaterra} \& {Ferrara}(2006)}]{SF06}
{Salvaterra}, R., \& {Ferrara}, A. 2006, \mnras, 367, L11

\bibitem[{{Salvaterra} {et~al.}(2010){Salvaterra}, {Ferrara}, \&
  {Dayal}}]{Salvaterra10}
{Salvaterra}, R., {Ferrara}, A., \& {Dayal}, P. 2010, ArXiv e-prints

\bibitem[{{Somerville} \& {Kolatt}(1999)}]{SK99}
{Somerville}, R.~S., \& {Kolatt}, T.~S. 1999, \mnras, 305, 1

\bibitem[{{Stark} {et~al.}(2007){Stark}, {Loeb}, \& {Ellis}}]{SLE07}
{Stark}, D.~P., {Loeb}, A., \& {Ellis}, R.~S. 2007, \apj, 668, 627

\bibitem[{{Trenti} \& {Stiavelli}(2008)}]{TS08}
{Trenti}, M., \& {Stiavelli}, M. 2008, \apj, 676, 767

\bibitem[{{Trenti} {et~al.}(2010){Trenti}, {Stiavelli}, {Bouwens}, {Oesch},
  {Shull}, {Illingworth}, {Bradley}, \& {Carollo}}]{Trenti10}
{Trenti}, M., {Stiavelli}, M., {Bouwens}, R.~J., {Oesch}, P., {Shull}, J.~M.,
  {Illingworth}, G.~D., {Bradley}, L.~D., \& {Carollo}, C.~M. 2010, \apjl, 714,
  L202

\bibitem[{{Vale} \& {Ostriker}(2004)}]{VO04}
{Vale}, A., \& {Ostriker}, J.~P. 2004, \mnras, 353, 189

\bibitem[{{Vale} \& {Ostriker}(2006)}]{VO06}
---. 2006, \mnras, 371, 1173

\bibitem[{{Vale} \& {Ostriker}(2008)}]{VO08}
---. 2008, \mnras, 383, 355

\bibitem[{{Wetzel} {et~al.}(2009){Wetzel}, {Cohn}, \& {White}}]{Wetzel09}
{Wetzel}, A.~R., {Cohn}, J.~D., \& {White}, M. 2009, \mnras, 395, 1376

\bibitem[{{Wetzel} \& {White}(2010)}]{WW10}
{Wetzel}, A.~R., \& {White}, M. 2010, \mnras, 403, 1072

\bibitem[{{Wyithe} \& {Loeb}(2006)}]{WL06}
{Wyithe}, J.~S.~B., \& {Loeb}, A. 2006, \nat, 441, 322

\end{thebibliography}
\end{document}